\documentclass[epj,twocolumn]{webofc}
\usepackage[varg]{txfonts}   % Web of Conferences font
\usepackage{epstopdf}
\usepackage{amssymb}
\usepackage{amsmath}

\woctitle{Flavour changing and conserving processes}
\begin{document}
\title{Leptoquarks in Flavour Physics}

\begin{minipage}[t]{\textwidth}
	\begin{flushright}
		PSI-PR-17-22\\
		ZU-TH 39/17
	\end{flushright}
\end{minipage}

\author{Dario M\"uller\inst{1,2}\fnsep\thanks{\email{dario.mueller@psi.ch}} 
}

\institute{Paul Scherrer Institut, CH--5232 Villigen PSI, Switzerland	
\and
 Physik Institut, Universit\"at Z\"urich, Winterthurstrasse 190,CH-8057 Z\"urich, Switzerland}   

\abstract{While the LHC has not directly observed any new particle so far, experimental results from LHCb, BELLE and BABAR point towards the violation of lepton flavour universality in $b\to s\ell^{+}\ell^{-}$ and $b\to c\ell\nu$. In this context, also the discrepancy in the anomalous magnetic moment of the muon can be interpreted as a sign of lepton flavour universality violation. Here we discuss how these hints for new physics can also be explained by introducing leptoquarks as an extension of the Standard Model. Indeed, leptoquarks are good candidates to explain the anomaly in the anomalous magnetic moment of the muon because of an $m_{t}/m_{\mu}$ enhanced contribution giving correlated effects in $Z$ boson decays which is particularly interesting in the light of future precision experiments.}
\maketitle
\section{Introduction}
\label{intro}
So far the LHC has not directly observed any particles beyond the ones in the Standard Model of particle physics (SM). While in the SM the Higgs boson couplings to fermions are the only source of lepton flavour universality violation (LFUV), in the past years several hints for additional LFUV have been accumulated. The ratios of semi-leptonic $B$ decays
\begin{align*}
R\left(K^{(*)}\right)&={\rm Br}\left[B\to K^{(*)}\mu^+\mu^-\right]{\bigg/}{\rm Br}\left[B\to K^{(*)}e^+e^-\right]\,,\\
R\left(D^{(*)}\right)&={\rm Br}\left[B\to D^{(*)}\tau\nu\right]{\bigg/}{\rm Br}\left[B\to D^{(*)}\ell\nu\right]\,,\\
R\bigg(J/\Psi\bigg)&={\rm Br}\bigg[B_c^{+}\to J/\Psi\tau^{+}\nu\bigg]{\bigg/}{\rm Br}\bigg[B_c\to J/\Psi\mu^{+}\nu\bigg]\,,
\end{align*}
show sizeable deviations from the SM predictions.\\
Considering $b\to s\ell^{+}\ell^{-}$ transitions, $R(K)$~\cite{Aaij:2014ora} and $R(K^{*})$~\cite{Aaij:2017vbb} show a combined significance for LFUV at the $4\,\sigma$ level~\cite{Altmannshofer:2017yso,DAmico:2017mtc,Hiller:2017bzc,Geng:2017svp,Hurth:2017hxg,Ciuchini:2017mik}. If we also take into account all other $b\to s\mu^+\mu^-$ observables, in several scenarios the global fit prefers new physics (NP) above the $5\,\sigma$~level \cite{Capdevila:2017bsm}.\\
In $R(D^{(*)})$ BaBar~\cite{Lees:2012xj}, BELLE~\cite{Huschle:2015rga,Hirose:2017dxl} and LHCb~\cite{Aaij:2015yra,Betti:2017xri} found a significance for LFUV of about $4\,\sigma$~\cite{Amhis:2016xyh}. In this context it is interesting to mention that LHCb recently measured a deviation of about $2\,\sigma$ in $R(J/\Psi)$ \cite{Aaij:2017tyk}. This is consistent with the anomaly in $R(D)$ and $R(D^{*})$.\\
The tension in the anomalous magnetic moment of the muon $(a_\mu)$ between measurement~\cite{Bennett:2006fi} and SM prediction is at the $3\,\sigma$ level~\cite{Nyffeler:2016gnb,Jegerlehner:2009ry}. Also this discrepancy can be interpreted as a sign of LFUV even though the anomalous magnetic moments are already in the SM flavour non-universal: If we assume that NP coupled with the same strength to electrons and to muons, $a_e$ would be more sensitive for NP. Since no deviation from the SM prediction in $a_e$ has been observed so far, the tension in $a_\mu$ can be interpreted as another sign for LFUV induced by NP.\\
Motivated by these anomalies, we study leptoquarks (LQ) which provide a possible solution of them. Firstly we concentrate on each anomaly itself and discuss which LQ representations suit best for an explanation and consider correlated effects in other observables. Finally, we will investigate if we can explain the described anomalies simultaneously.

\begin{figure*}[th]
	\begin{center}
		\begin{tabular}{cp{7mm}c}
			\includegraphics[width=0.35\textwidth]{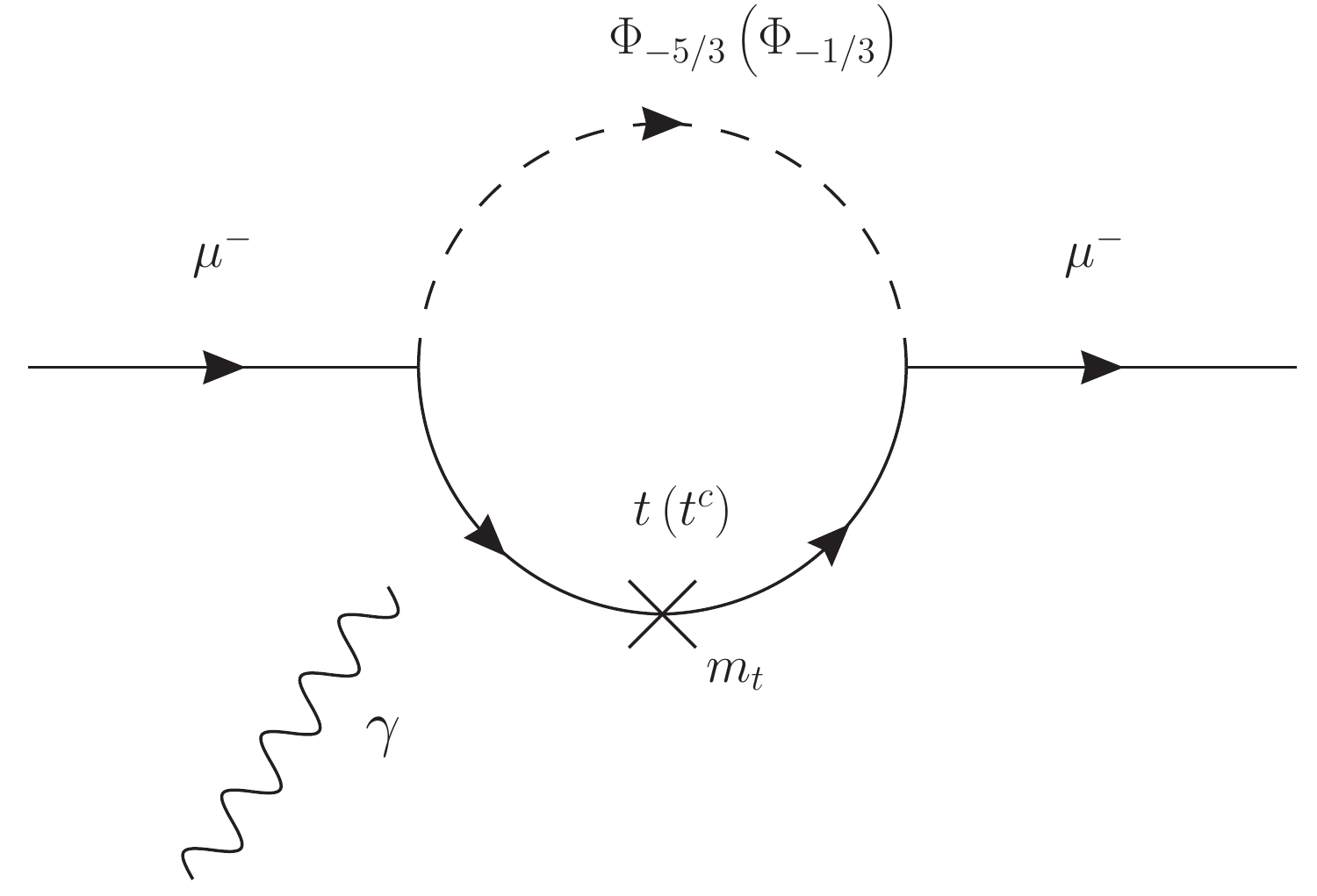}
			\includegraphics[width=0.35\textwidth]{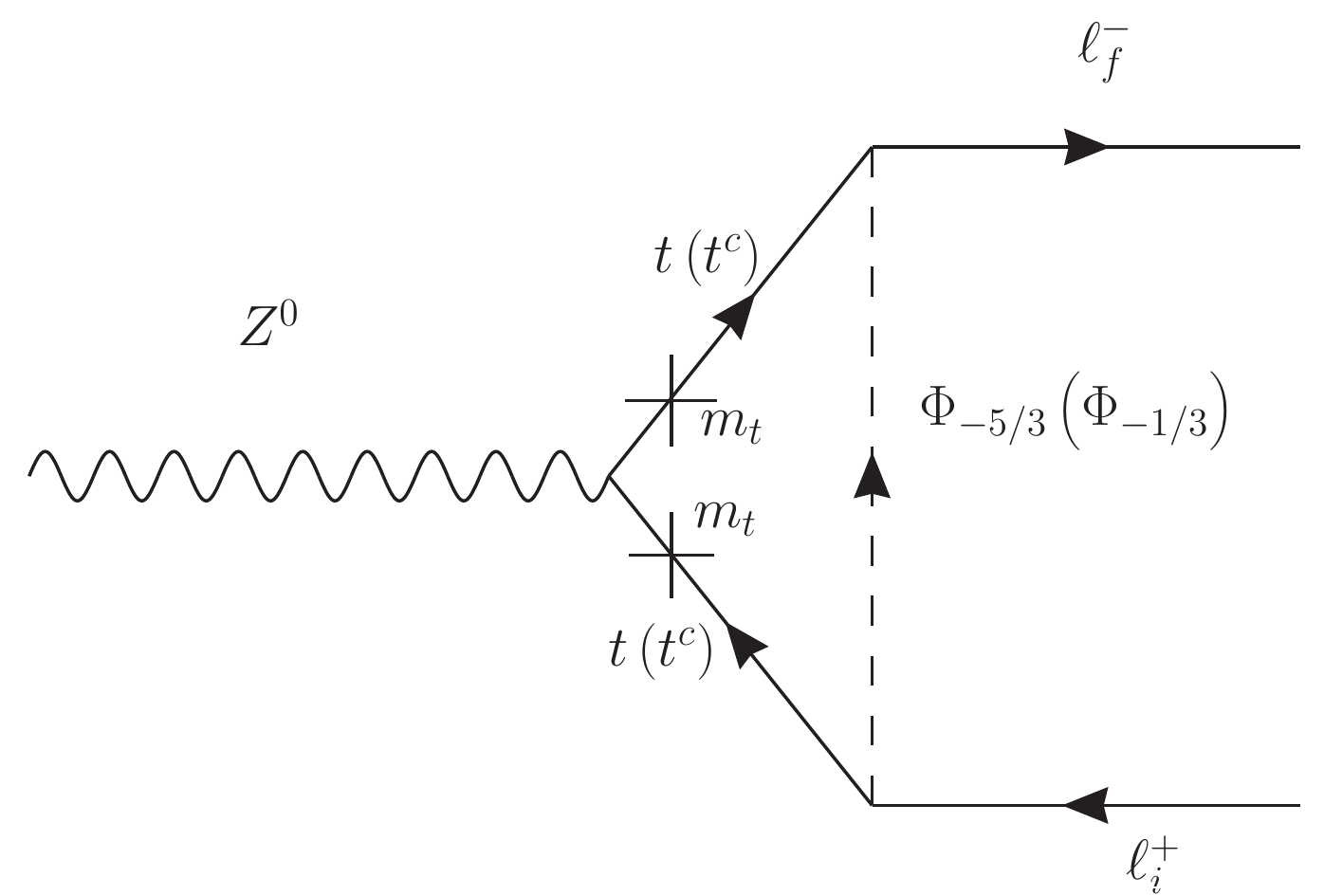}
		\end{tabular}
	\end{center}
	\caption{Feynman diagrams that contribute to $a_{\mu}$ and $Z\to\ell_{i}^{+}\ell_{f}^{-}$. The crosses on the fermion lines stand for chirality flips, involving a top mass.}
	\label{Fig:FeynDiag}
\end{figure*}

\section{$a_\mu$}\label{Sec:AMM}
\begin{figure*}[tb]
	\begin{center}
		\begin{tabular}{cp{7mm}c}
			\includegraphics[width=0.59\textwidth]{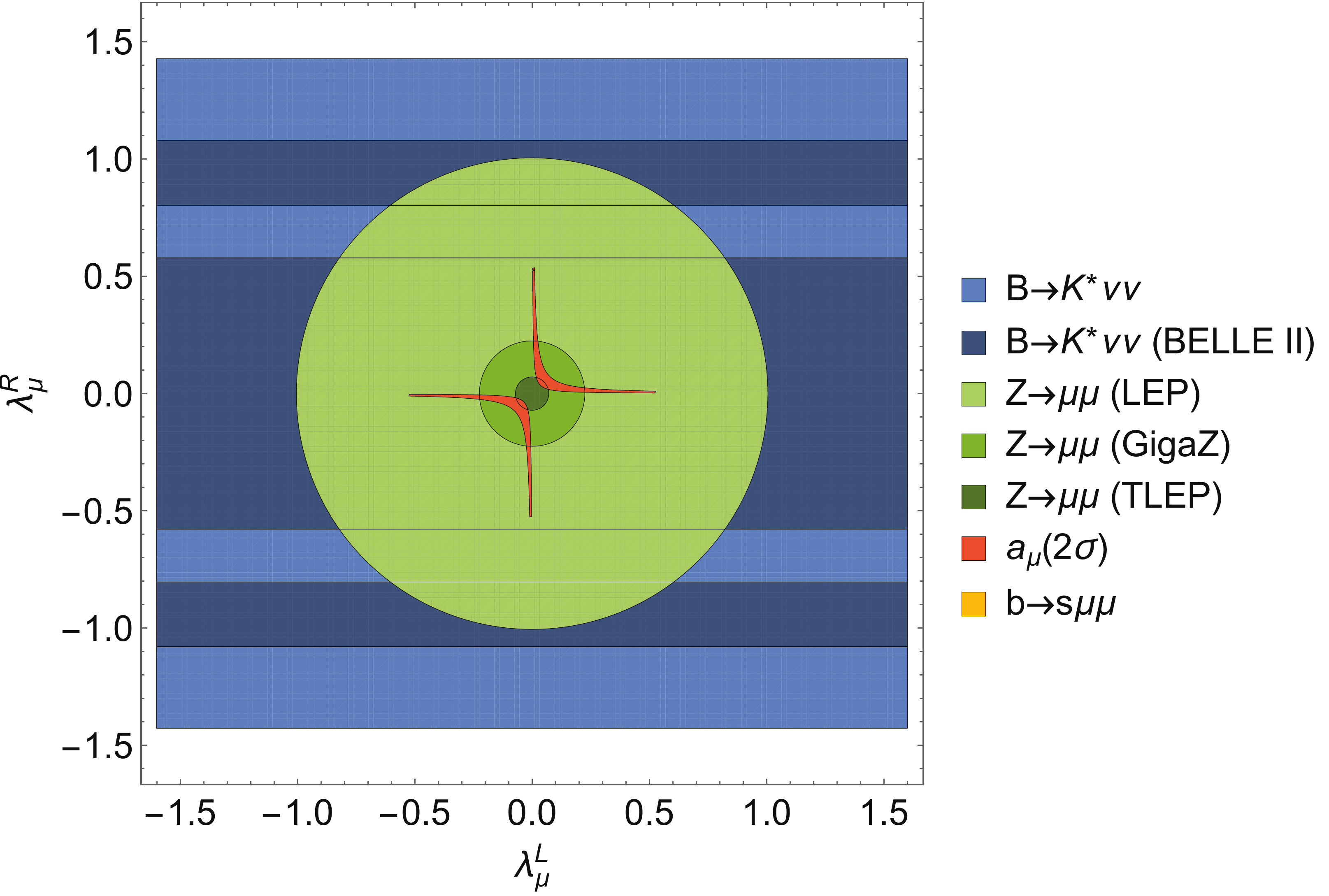}
			\includegraphics[width=0.41\textwidth]{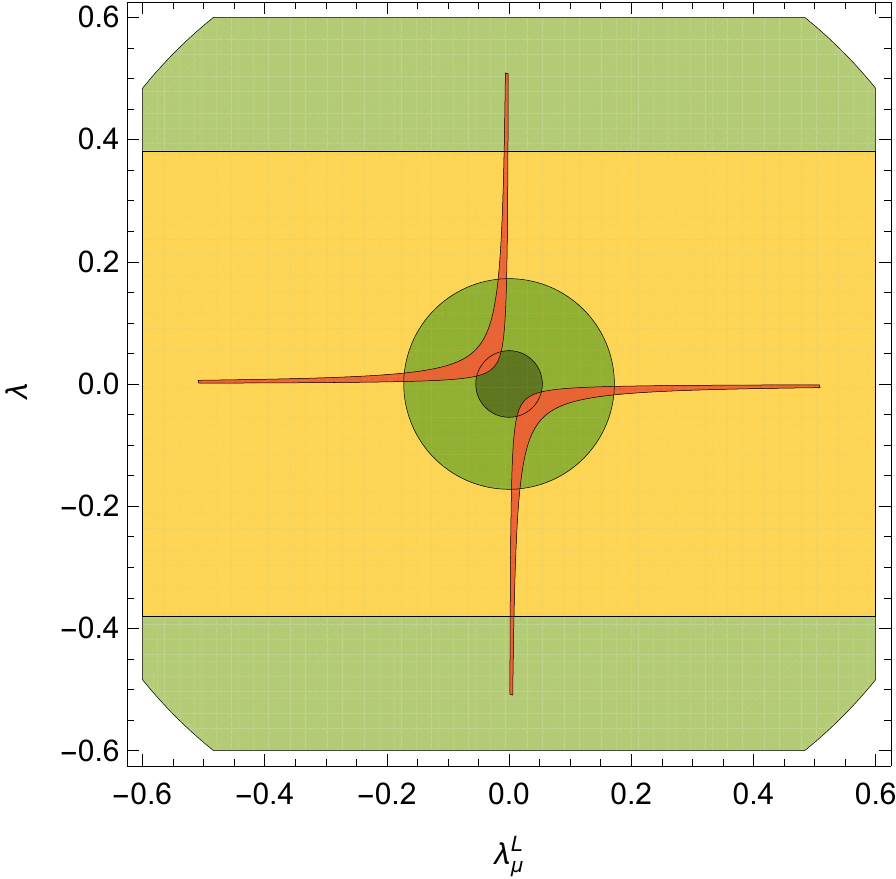}
		\end{tabular}
	\end{center}
	\caption{Allowed regions in the $\lambda^L_\mu$-$\lambda^R_\mu$ parameter space from current limits of LEP and expected future constraints of GigaZ and TLEP with $M_{LQ}=1~\mathrm{TeV}$. Left: $\mathrm{SU}(2)$ singlet $\Phi_1$. Right: $\mathrm{SU}(2)$ doublet $\Phi_2$. Note that if the expected sensitivity of TLEP was reached and no excess in the $Z\mu\bar{\mu}$-coupling was found, an explanation of the anomaly in $a_{\mu}$ would strongly be disfavoured.}
	\label{muLmuR}
\end{figure*}

In order to explain the descrepancy in $a_{\mu}$, a rather large effect of the order of the SM weak interaction contribution is required. Among the set of new particles that can give the desired effect, LQ are interesting candidates \cite{Djouadi:1989md,Davidson:1993qk,Couture:1995he,Bauer:2015knc,Das:2016vkr}. Even though the LQ must be rather heavy due to LHC constraints (above 1 TeV), one can still get relevant effects in $a_{\mu}$. For some representations of LQ, the amplitude can be enhanced by a factor $m_t/m_\mu\simeq\mathcal{O}(10^{3})$ or $m_{b}/m_{\mu}\simeq\mathcal{O}(10)$ compared to the SM.\\
To achieve an enhancement of factor $m_t/m_\mu$, the LQ must couple to left- and right-handed top or anti-top quarks simultaneously. However, not all LQ representations have this feature. Among the 10 scalar and vector LQ representations generating lepton-quark interaction terms that are invariant under the SM gauge group~\cite{Buchmuller:1986zs}, only two scalars can possess this enhancement: An $\mathrm{SU}(2)$ singlet $\Phi_{1}$ and an $\mathrm{SU}(2)$ doublet $\Phi_{2}$ with quantum numbers
\begin{align}
Q\left( {\Phi _1^{}} \right):\left( {3,1, - \frac{2}{3}} \right)\,,
~~~~Q\left( {{\Phi _2}} \right):\left( {\bar 3,2, - \frac{7}{3}} \right)\,.
\end{align}
These representations couple to the fermions of the SM as follows
\begin{align}
\mathcal{L}_{int}=&\left(\lambda_{fi}^{1R}\overline{u^{c}_{f}}\ell_{i}+\lambda_{fi}^{1L}\overline{Q^{c}_{f}}i\tau_{2}L_{i}\right)\Phi^{\dagger}_{1}\,,\label{eq:Phi1}\\
\mathcal{L}_{int}=&\left(\lambda_{fi}^{2RL}\overline{u_f}L_{i}+\lambda_{fi}^{2LR}\overline{Q_f}i\tau_2\ell_i\right)\Phi_{2}^{\dagger}\,.
\end{align}
$Q$ and $L$ are $\mathrm{SU}(2)$ doublets, $u$ and $\ell$ are $\mathrm{SU}(2)$ singlets and $c$ denotes charge conjugation. $f$ and $i$ are flavour indices. The Feynman diagrams involving $\Phi_1$ and $\Phi_2$ are shown left in Fig. \ref{Fig:FeynDiag}.\\
For simplicity, we only consider couplings of top quarks to muons and set $\lambda_{32}^{1L,1R}$ and $\lambda_{32}^{2RL,2LR}$ equal to $\lambda_{\mu}^{L,R}$. Besides the contribution to $a_{\mu}$, we also obtain correlated effects in $Z\to\mu^+\mu^-$ (see the right Feynman diagram in Fig. \ref{Fig:FeynDiag}), in $b\to s\nu\bar\nu$ (for $\Phi_1$) and in $b\to s\mu^{+}\mu^{-}$ (for $\Phi_{2}$). Fig. \ref{muLmuR} shows the allowed regions in $\lambda_{\mu}^{L}$-$\lambda_{\mu}^{R}$ parameter space for a LQ with mass of $1~\mathrm{TeV}$, to respect constraints from direct searches. For $Z\to\mu^+\mu^-$ the expected bounds from GigaZ \cite{Baer:2013cma} and TLEP \cite{Gomez-Ceballos:2013zzn} are shown as well. Additionally, for $\Phi_{1}$ the expected BELLE~II limit for $B\to K^{(*)}\nu\bar{\nu}$ is taken into account. $\Phi_{2}$ gives a $C_{9}=C_{10}$-like contribution to $b\to s\ell^{+}\ell^{-}$ which provides a slight improvement of $1\,\sigma$ in the global fit.\\
If we also allow for couplings of the top quark to taus or electrons, lepton flavour violating processes as $\tau\to\mu\gamma$ or $Z\to\mu e$ will be possible. A more detailed discussion can be found in Ref. \cite{ColuccioLeskow:2016dox}.

\section{$b\to s\ell^+\ell^-$}\label{sec:bsellell}

\begin{figure*}
	\begin{center}
		\includegraphics[width=15cm]{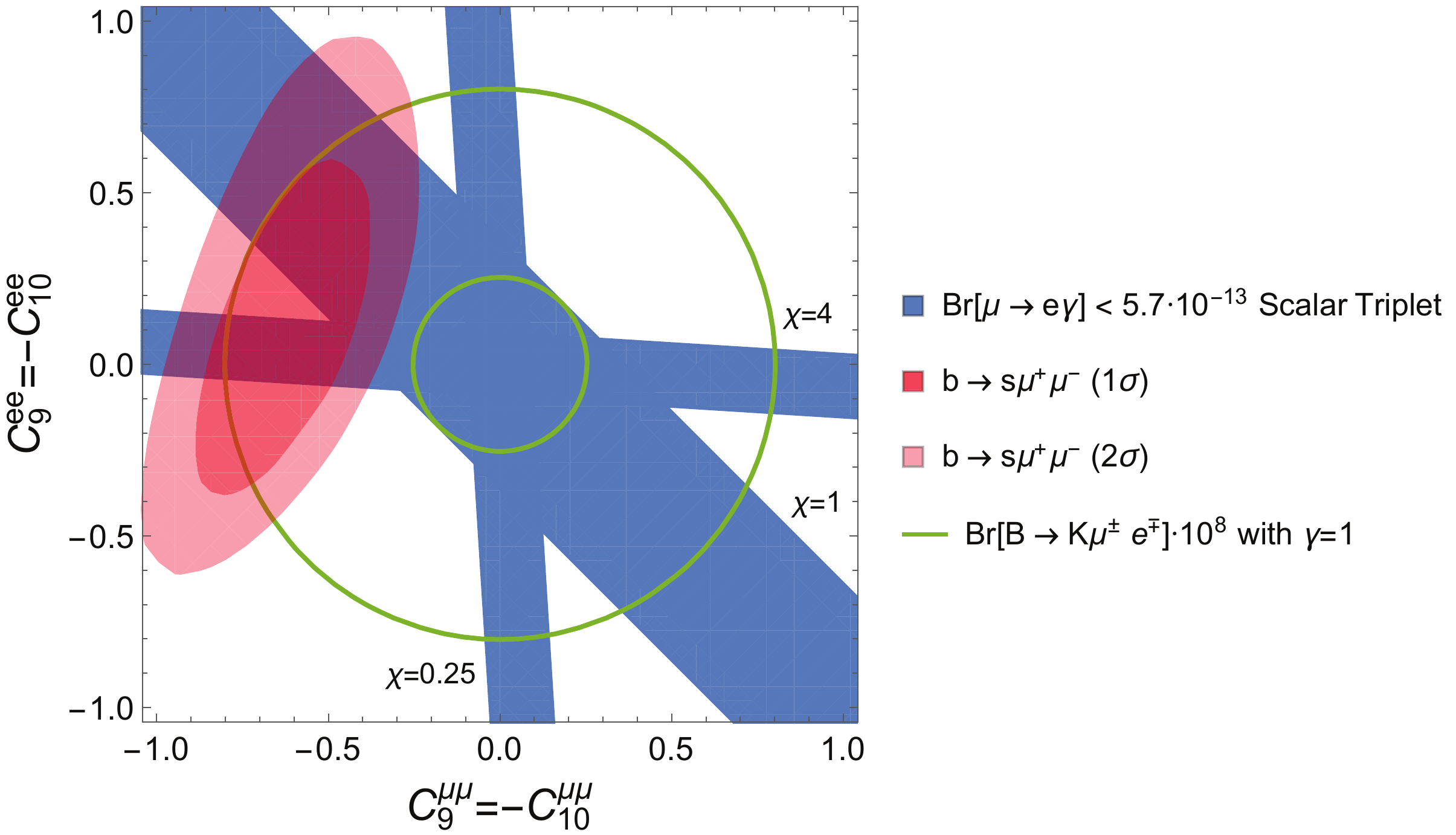}
	\end{center}
	\caption{The darker (lighter) red region shows the global fit at $1\,\sigma$ ($2\,\sigma$) for $b\to se^{+}e^{-}$ and $b\to s\mu^{+}\mu^{-}$ transitions. In blue, the allowed region constrained by the current limit $\rm{Br}\left[\mu\to e \gamma\right]<4.2\times 10^{-13}$ \cite{TheMEG:2016wtm} with $\chi=\{1/4,1,4\}$ for the scalar triplet $\Phi_{3}$ with $M_{LQ}=1\mathrm{TeV}$. The outer (inner) green contour shows the case where $\rm{Br}\left[B\to K\mu e\right]=2\times10^{-8}(0.2\times10^{-8})$. For real values of $\chi$ only the region with $C_9^{\mu\mu}<0$ and $C_9^{ee}>0$ seems to be compatible with experimental limits. Indeed this is also where the central value of the global fit lies. However, if in the future the global fit prefers $C_{9}^{\mu\mu}<0$ and $C_{9}^{ee}<0$, LQ with real couplings will not be able to explain $b\to s\ell^{+}\ell^{-}$ transitions without violating the constraints from $\mu\to e\gamma$.}
	\label{fig:C9eeC9mumu}
\end{figure*}

Dealing with $b\to s\ell^{+}\ell^{-}$ transitions, one usually works with an effective Hamiltonian of the form
\begin{align}
H_{\rm eff}^{\ell_f\ell_i}=- \dfrac{ 4 G_F }{\sqrt 2}V_{tb}V_{ts}^{*} \sum\limits_{a = 9,10} C_a^{fi} O_a^{fi}+C_a^{\prime fi} O_a^{\prime fi}\,,\\
{O_{9(10)}^{fi}} =\dfrac{\alpha }{4\pi}[\bar s{\gamma ^\mu } P_L b]\,[\bar\ell_f{\gamma _\mu }(\gamma_5)\ell_i] \,,\\
{O_{9(10)}^{\prime fi}} =\dfrac{\alpha }{4\pi}[\bar s{\gamma ^\mu } P_R b]\,[\bar\ell_f{\gamma _\mu }(\gamma_5)\ell_i] \,.
\label{eq:effHam}
\end{align}
Then, for several scenarios with NP in $C_9$ only, $C_9=-C_{10}$ or $C_9=-C_{9}^\prime$, the global fit prefers NP above the $5\,\sigma$ level~\cite{Capdevila:2017bsm}.\\
We consider the case where NP contributes as $C_{9}=-C_{10}$. While in the muon channel the preferred value deviates from the SM prediction with a tension of $5\,\sigma$, in the case with electrons the SM contribution is sufficient to give a good fit to the data. To reach the central value, an $\mathcal{O}(10\%)$ NP effect in the muon channel is required. $b\to s\ell^+\ell^-$ is suppressed by a loop and a CKM factor because it is a flavour changing neutral current. Therefore, the effect of NP does not have to be very large to account for the data. There are several suggestions to explain the anomaly in $b\to s\mu^{+}\mu^{-}$ with LQ, see e.g. \cite{Gripaios:2014tna,Fajfer:2015ycq,Becirevic:2015asa,Greljo:2015mma,Calibbi:2015kma,Alonso:2015sja,Barbieri:2015yvd,Becirevic:2016yqi,Becirevic:2016oho,Becirevic:2017jtw,Aloni:2017ixa,Chauhan:2017ndd,Sumensari:2017mud,Chen:2017hir}.\\
There are three LQ representations that give a $C_{9}=-C_{10}$-like contribution to $b\to s\ell^{+}\ell^{-}$
\begin{align}
\Phi_{3}:~&\left(3,3,-\frac{2}{3}\right)\,,~\mathcal{L}_{int}=\lambda_{fi}^{3}\overline{Q^{c}_{f}}i\tau_{2}\left(\tau\cdot\Phi_{3}\right)^{\dagger}L_{i}\label{eq:Phi3}\,,\\
V^{\mu}_{1}:~&\left(\bar{3},1,-\frac{4}{3}\right)\,,~\mathcal{L}_{int}=\left(\kappa_{fi}^{1R}\overline{d_f}\gamma_{\mu}\ell_{i}+\kappa_{fi}^{1L}\overline{Q_f}\gamma_{\mu}L_{i}\right)V_{1}^{\mu\dagger}\label{eq:vecsing}\,,\\
V^{\mu}_{3}:~&\left(3,3,\frac{4}{3}\right)\,,~\mathcal{L}_{int}=\kappa_{fi}^{3}\overline{Q_f}\gamma_{\mu}\left(\tau\cdot V_{3}^{\mu}\right)L_{i}\,.
\end{align}
If we only allow for couplings to muons, these LQ will give tree-level effects in $b\to s\mu^{+}\mu^{-}$ but contribute at loop-level in other flavour observables. This is why LQ can explain the anomalies in $b\to s\mu^{+}\mu^{-}$ transitions but are not in conflict with other observables.\\ 
Including couplings to electrons, we get additional effects in $b\to se^{+}e^{-}$ as well as in the lepton flavour violating processes $\mu\to e\gamma$ and $b\to s \mu e$. We can write
\begin{align}
{\rm Br}\left[\mu\to e \gamma\right]&\propto\left|~\chi C_{9}^{ee}+\frac{C_{9}^{\mu\mu}}{\chi}\right|^2\,,\\
{\rm Br}\left[B\to K\mu e\right]&\propto\left|\frac{C_{9}^{ee}}{\gamma}\right|^{2}+\left|\gamma\,C_{9}^{\mu\mu}\right|^2\,,
\end{align}
where $\chi=y_{32}/y_{21}$ and $\gamma=y_{21}/y_{22}$, with $y=\lambda$ for scalar LQ and $y=\kappa$ for vector LQ. Note that one can avoid an effect in $\mu\to e\gamma$ completely if $C_{9}^{\mu\mu}=-\chi^2C_{9}^{ee}$. For real values of $\chi$ this means that ${\rm sign}[C_{9}^{ee}]=-{\rm sign}[C_{9}^{\mu\mu}]$. In Fig. \ref{fig:C9eeC9mumu} you find a detailed analysis for the scalar triplet $\Phi_{3}$. For $V_{1}^{\mu}$ and $V_{3}^{\mu}$ the situation is quite similar, albeit the blue bands are more narrow.  For more information see Ref. \cite{Crivellin:2017dsk}.

\section{$b\to c\tau\nu$} \label{sec:RD}
\begin{figure}
	\centering
	\includegraphics[width=8cm,clip]{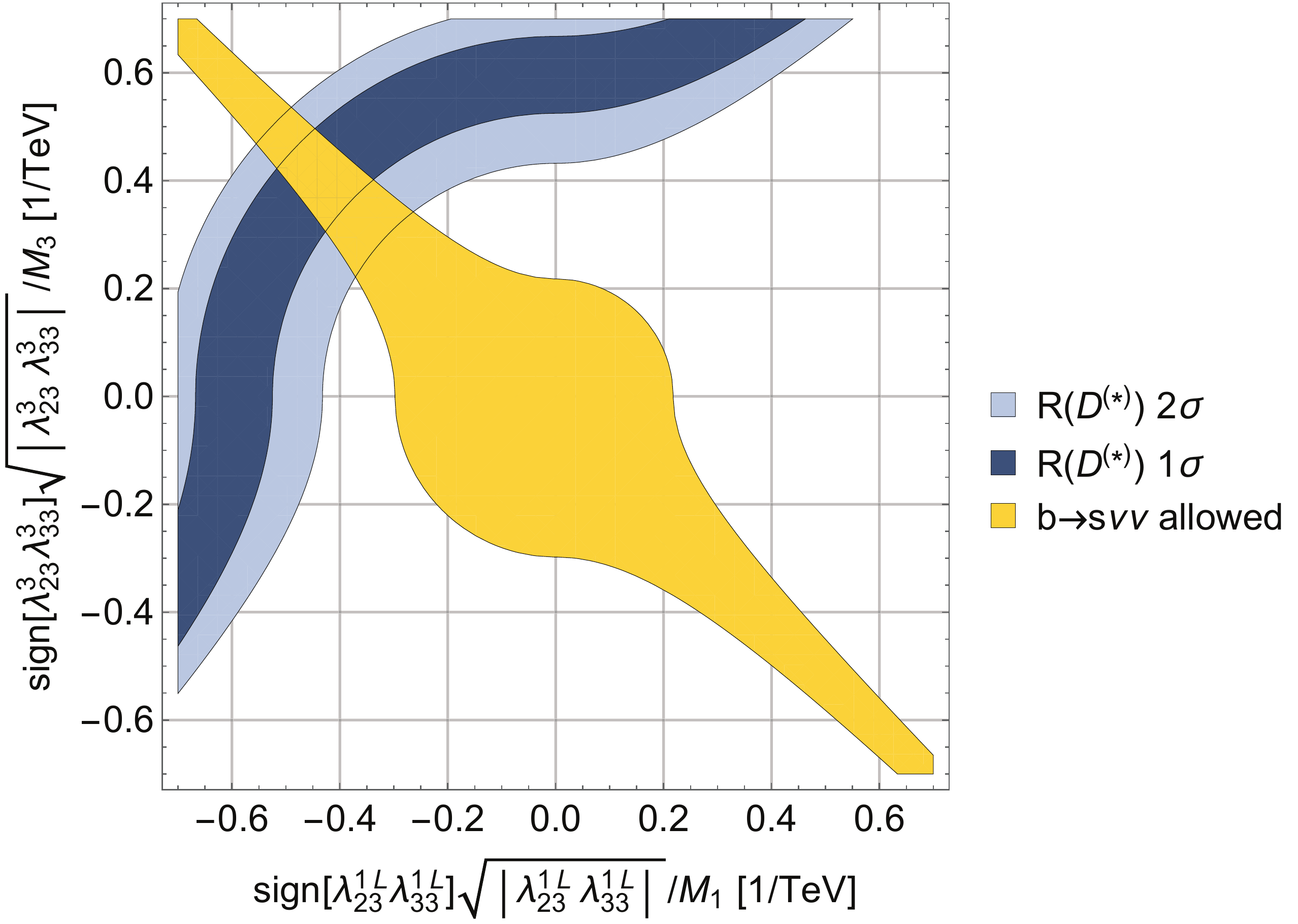}
	\caption[width=\textwidth]{In lighter (darker) blue you see the $1\,\sigma$ ($2\,\sigma$) preferred region of the weighted sum of $R(D)$ and $R(D^{*})$, in yellow the allowed region of $b\to s\nu\bar{\nu}$. $M_{1(3)}$ is the mass of the LQ $\Phi_{1}$ ($\Phi_{3}$).}
	\label{Fig:kappa}
\end{figure}

\begin{figure}
	\centering
	\includegraphics[width=8cm,clip]{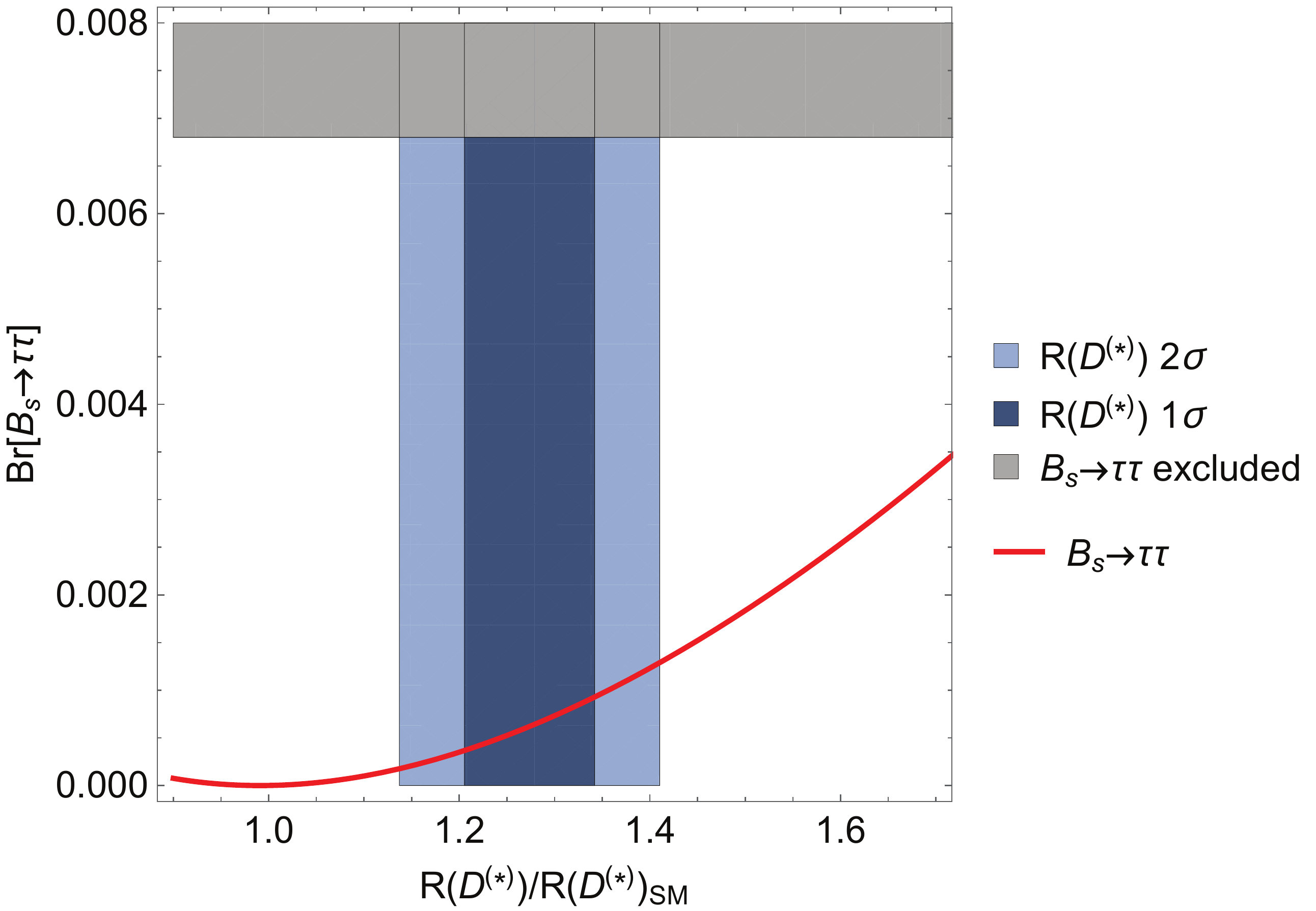}
	\caption{In blue we have the weighted sum of $R(D)$ and $R(D^{*})$ and the red line depicts the correlated prediction of ${\rm{Br}\left[B_{s}\to\tau^{+}\tau^{-}\right]}$. The grey region shows the current experimental bounds.}
	\label{Fig:Bstautau}
\end{figure}

The transition $b\to c\tau\nu$ is a charged current process and is therefore generated at tree-level in the SM. Since an explanation of $R(D)$ and $R(D^{*})$ requires an $\mathcal{O}(10\%)$ effect, also the NP must contribute at tree-level. Charged Higgses \cite{Crivellin:2012ye,Tanaka:2012nw,Celis:2012dk,Crivellin:2013wna,Crivellin:2015hha}, $W^{\prime}$ bosons \cite{Greljo:2015mma} and LQ \cite{Alonso:2015sja,Calibbi:2015kma,Bauer:2015knc,Becirevic:2016yqi,Barbieri:2016las,Chauhan:2017uil,Dorsner:2017ufx,Popov:2016fzr,Hiller:2016kry,Sahoo:2016pet,Das:2016vkr,Altmannshofer:2017poe} are NP candidates. However, $B_c$ lifetime~\cite{Akeroyd:2017mhr,Freytsis:2015qca,Celis:2016azn,Li:2016vvp} and $q^2$ distributions~\cite{Celis:2016azn} in combination with direct searches \cite{Faroughy:2016osc} strongly disfavour charged Higgses and $W^{\prime}$ bosons.\\
Hence, we are left with LQ. If we want to explain the excess in $R(D)$ and $R(D^{*})$ by modifying the couplings to tau neutrinos and/or taus, we also get effects in $b\to s\tau^{+}\tau^{-}$ and $b\to s\nu\bar{\nu}$. It can be shown in a model independent way that these effects are of the order of $10^{3}$ compared to the SM \cite{Alonso:2015sja,Capdevila:2017iqn}. This is due to the fact that the processes $b\to s\tau^{+}\tau^{-}$ and $b\to s\nu\bar{\nu}$ occur at loop-level in the SM, while LQ contribute at tree-level with considerably large couplings in order to explain $R(D^{(*)})$. An $\mathcal{O}(10^{3})$ effect is in strong conflict with experimental results of $B\to K^{(*)}\nu\bar{\nu}$.\\
Hence, a model that explains $R(D)$ and $R(D^{*})$ must avoid a major effect in $b\to s\nu\bar{\nu}$. One possibility is the vector LQ singlet $V_{1}^{\mu}$ (see Eq. \eqref{eq:vecsing}). Another model was proposed in Ref. \cite{Crivellin:2017zlb}, where instead of $V_{1}^{\mu}$ two scalar LQ (the $\mathrm{SU}(2)$ singlet $\Phi_{1}$ in Eq. \eqref{eq:Phi1} and the $\mathrm{SU}(2)$ triplet $\Phi_{3}$ in Eq. \eqref{eq:Phi3}) are required. If Fig. \ref{Fig:kappa} we can see for which parameters we can explain $R(D^{(*)})$ without violating the limits of $b\to s\nu\bar{\nu}$.\\
Besides explaining $R(D^{(*)})$, it also enhances $B_{s}\to\tau^{+}\tau^{-}$ up to a factor $10^3$ compared to the SM and therefore lies within experimental reach. The current limit is ${\rm Br}\left[B_{s}\to\tau^{+}\tau^{-}\right]<6.8\times 10^{-3}$ \cite{Aaij:2017xqt}. Fig. \ref{Fig:Bstautau} shows the correlation of $R(D^{(*)})$ and ${\rm Br}\left[B_{s}\to\tau^{+}\tau^{-}\right]$.

\section{Simultaneous Explanation}
\begin{figure*}[tb]
	\begin{center}
		\begin{tabular}{cp{7mm}c}
			\includegraphics[width=0.55\textwidth]{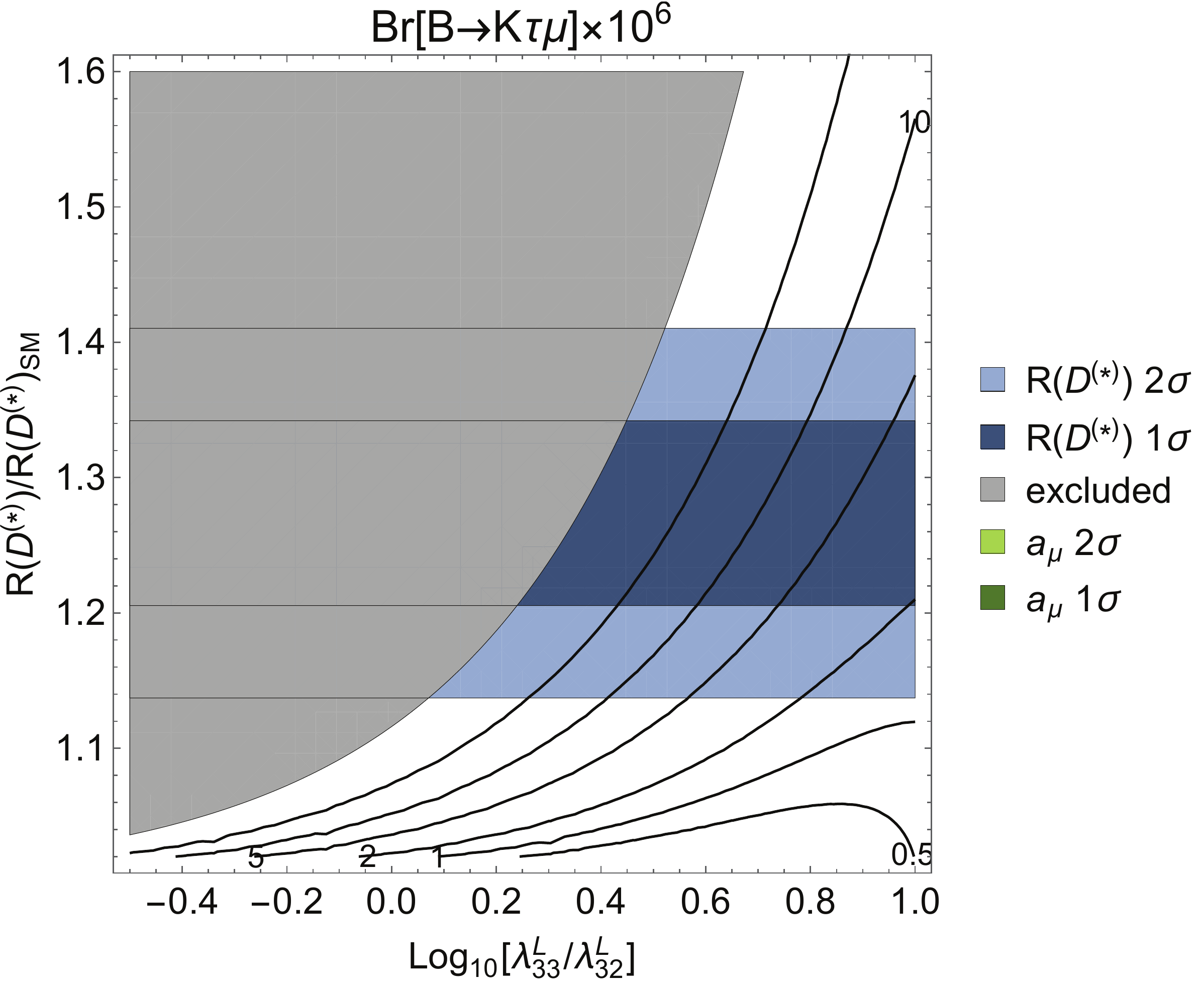}
			\includegraphics[width=0.44\textwidth]{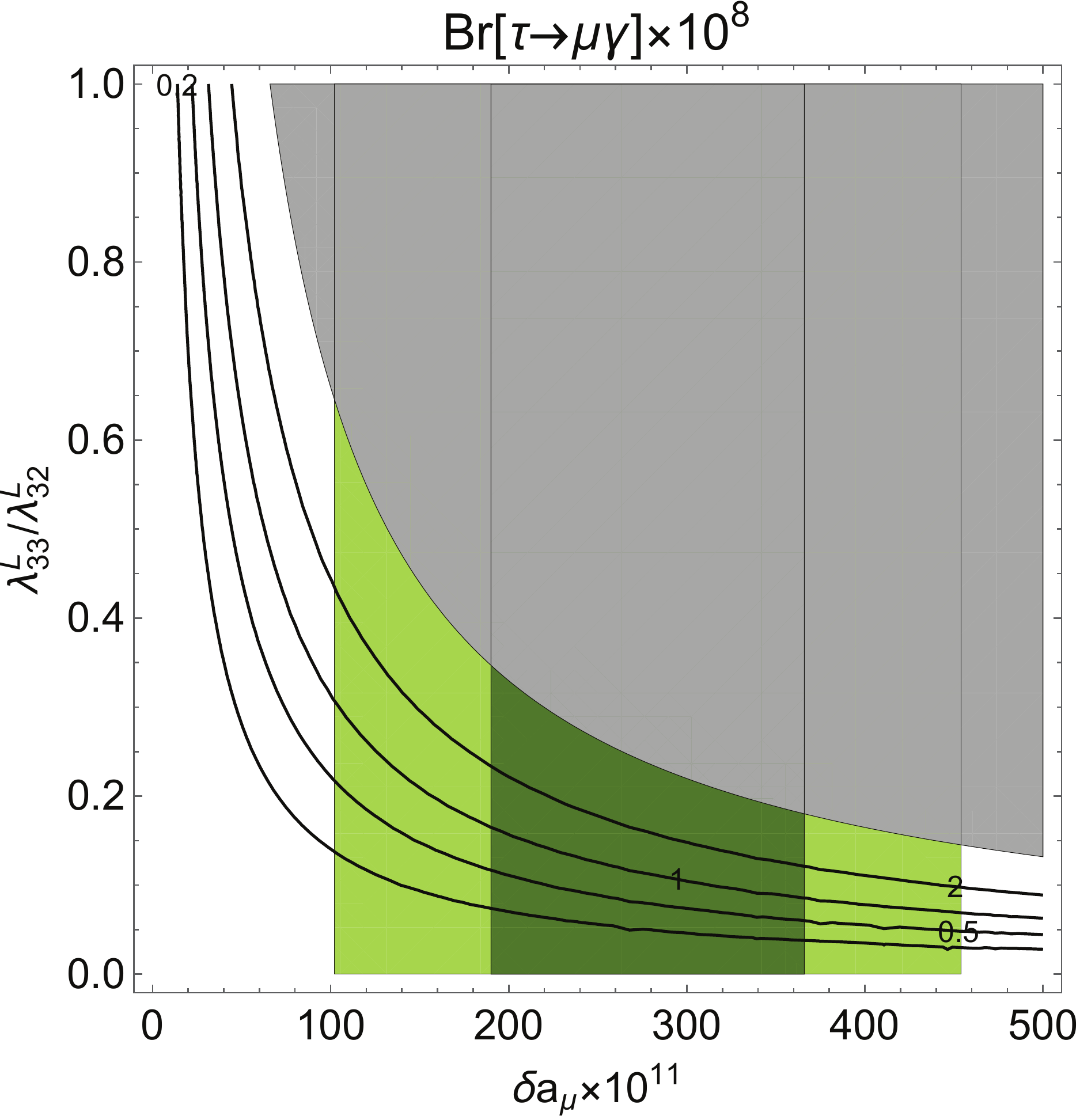}
		\end{tabular}
	\end{center}
	\caption{Left: In blue we have the weighted sum of $R(D)$ and $R(D^{*})$. The contour lines show ${\rm Br}\left[B\to K\tau\mu\right]\times 10^{6}$. Right: The anomaly in $a_{\mu}$ combined with the contour lines depicting ${\rm Br}\left[\tau\to\mu\gamma\right]\times 10^8$. In both plots the grey region is excluded by experiments.}
	\label{Fig:BK-AMM}
\end{figure*}

We may ask if one can explain the three types of anomalies in $a_{\mu}$, $b\to s\mu^{+}\mu^{-}$ and $R(D^{(*)})$ simultaneously.\\
If we take the model that we proposed in Sec. \ref{sec:RD} and insert a discrete symmetry for the couplings $\lambda _{jk}^{1L}=e^{i\pi j}\lambda_{jk}^{3}\equiv \lambda _{jk}^{L}$ the effect in $b\to s\nu\bar{\nu}$ will be cancelled exactly.
If we include $b\to s\mu^{+}\mu^{-}$, $B\to D^{(*)}\mu\nu/B\to D^{(*)}e\nu$ will be affected. However, these observables remain compatible with experiments. Additionally the LFV process $\tau\to\mu\gamma$ and $b\to s\tau\mu$ appear. Both processes give constraints on our parameter space. On the left-hand side of Fig. \ref{Fig:BK-AMM} the allowed regions for $M_{LQ}=1~\mathrm{TeV}$ are shown, under the assumption that $b\to s\mu^{+}\mu^{-}$ matches the central value of the global fit. As we can see easily, the explanation of $R(D^{(*)})$ requires $\lambda^{L}_{33}/\lambda^{L}_{32}>1$. On the right-hand side of Fig. \ref{Fig:BK-AMM} we need $\lambda^{L}_{33}/\lambda^{L}_{32}<0.6$ to explain $a_{\mu}$ without getting in conflict with experimental constraints of $\tau\to\mu\gamma$. With this contradiction, we cannot explain all three anomalies simultaneously. However, if we abandon anyone of them, it is possible to explain the other two anomalies. If we abandon the assumption $\lambda _{jk}^{1L}=e^{i\pi j}\lambda_{jk}^{3}$ one can avoid the discussed constraints and explain all three anomalies.\\
Another good candidate is the vector LQ singlet $V_{1}^{\mu}$ because it does not affect the $b\to s\nu\bar{\nu}$ transitions but gives a $C_{9}=-C_{10}$ effect in $b\to s\ell^{+}\ell^{-}$. It can also be used to explain $R(D^{(*)})$. In $a_{\mu}$ we can obtain an $m_{b}/m_{\mu}$ contribution which is not as large as for the two scalar LQ $\Phi_{1}$ and $\Phi_{2}$ (see Sec.\ref{Sec:AMM}), but still may be significant.\\
However, one faces the problem that massive vector bosons without a Higgs mechanism are not renormalizable. For $\ell^{\prime}\to\ell\gamma$ in Sec. \ref{sec:bsellell} this does not cause a problem because this process is finite in unitary gauge. Recently, models containing renormalizable massive vector LQ were proposed \cite{DiLuzio:2017vat,Calibbi:2017qbu,Bordone:2017bld,Barbieri:2017tuq}. In Ref. \cite{Calibbi:2017qbu} a Pati-Salam model in combination with vector-like fermions is phenomenologically consistent and leads to a massive vector LQ. This vector LQ has the properties of $V_{1}^{\mu}$ in Eq. \eqref{eq:vecsing}.

\section{Conclusion}
In these proceedings we reviewed which LQ are good candidates to explain flavour anomalies: For the deviation in $a_{\mu}$ the scalar LQ $\Phi_{1}$ and $\Phi_{2}$ give an $m_{t}/m_{\mu}$ enhanced contribution. This is a useful feature to explain the rather large discrepancy. Considering $b\to s\ell^{+}\ell^{-}$ transitions, the three LQ representations $\Phi_{3}$, $V_{1}^{\mu}$ and $V_{3}^{\mu}$ are required by the fit, because they give $C_{9}=-C_{10}$. $R(D^{(*)})$ can be explained either by $V_{1}^{\mu}$ or, as we proposed, by a combination of a scalar LQ singlet $\Phi_{1}$ and scalar LQ triplet $\Phi_{3}$ without getting in conflict with $b\to s\nu\bar{\nu}$ observables. We also showed that this model is able to explain any two of the three anomalies ($a_{\mu}$, $b\to s\mu^{+}\mu^{-}$ and $R(D^{(*)}$) simultaneously.

\subsection*{Acknowledgments}
D.M. thanks the organizing committee, and especially Giancarlo D'Ambrosio, for the invitation and for giving the opportunity to present his work.\\
D.M. is supported by an Ambizione Grant of the Swiss National Science Foundation (PZ00P2\_154834).

%
% BibTeX or Biber users please use (the style is already called in the class, ensure that the "woc.bst" style is in your local directory)
% \bibliography{name or your bibliography database}

\begin{thebibliography}{}
%\cite{Aaij:2014ora}
\bibitem{Aaij:2014ora}
R.~Aaij {\it et al.} [LHCb Collaboration],
%``Test of lepton universality using $B^{+}\rightarrow K^{+}\ell^{+}\ell^{-}$ decays,''
Phys.\ Rev.\ Lett.\  {\bf 113} (2014) 151601
doi:10.1103/PhysRevLett.113.151601
[arXiv:1406.6482 [hep-ex]].
%%CITATION = doi:10.1103/PhysRevLett.113.151601;%%
%501 citations counted in INSPIRE as of 02 Feb 2018


%\cite{Aaij:2017vbb}
\bibitem{Aaij:2017vbb}
R.~Aaij {\it et al.} [LHCb Collaboration],
%``Test of lepton universality with $B^{0} \rightarrow K^{*0}\ell^{+}\ell^{-}$ decays,''
JHEP {\bf 1708} (2017) 055
doi:10.1007/JHEP08(2017)055
[arXiv:1705.05802 [hep-ex]].
%%CITATION = doi:10.1007/JHEP08(2017)055;%%
%122 citations counted in INSPIRE as of 02 Feb 2018


%\cite{Altmannshofer:2017yso}
\bibitem{Altmannshofer:2017yso}
W.~Altmannshofer, P.~Stangl and D.~M.~Straub,
%``Interpreting Hints for Lepton Flavor Universality Violation,''
Phys.\ Rev.\ D {\bf 96} (2017) no.5,  055008
doi:10.1103/PhysRevD.96.055008
[arXiv:1704.05435 [hep-ph]].
%%CITATION = doi:10.1103/PhysRevD.96.055008;%%
%79 citations counted in INSPIRE as of 02 Feb 2018


%\cite{DAmico:2017mtc}
\bibitem{DAmico:2017mtc}
G.~D'Amico, M.~Nardecchia, P.~Panci, F.~Sannino, A.~Strumia, R.~Torre and A.~Urbano,
%``Flavour anomalies after the $R_{K^*}$ measurement,''
JHEP {\bf 1709} (2017) 010
doi:10.1007/JHEP09(2017)010
[arXiv:1704.05438 [hep-ph]].
%%CITATION = doi:10.1007/JHEP09(2017)010;%%
%79 citations counted in INSPIRE as of 02 Feb 2018


%\cite{Hiller:2017bzc}
\bibitem{Hiller:2017bzc}
G.~Hiller and I.~Nisandzic,
%``$R_K$ and $R_{K^{\ast}}$ beyond the standard model,''
Phys.\ Rev.\ D {\bf 96} (2017) no.3,  035003
doi:10.1103/PhysRevD.96.035003
[arXiv:1704.05444 [hep-ph]].
%%CITATION = doi:10.1103/PhysRevD.96.035003;%%
%64 citations counted in INSPIRE as of 02 Feb 2018


%\cite{Geng:2017svp}
\bibitem{Geng:2017svp}
L.~S.~Geng, B.~Grinstein, S.~Jäger, J.~Martin Camalich, X.~L.~Ren and R.~X.~Shi,
%``Towards the discovery of new physics with lepton-universality ratios of $b\to s\ell\ell$ decays,''
Phys.\ Rev.\ D {\bf 96} (2017) no.9,  093006
doi:10.1103/PhysRevD.96.093006
[arXiv:1704.05446 [hep-ph]].
%%CITATION = doi:10.1103/PhysRevD.96.093006;%%
%73 citations counted in INSPIRE as of 02 Feb 2018


%\cite{Hurth:2017hxg}
\bibitem{Hurth:2017hxg}
T.~Hurth, F.~Mahmoudi, D.~Martinez Santos and S.~Neshatpour,
%``Lepton nonuniversality in exclusive $b{\rightarrow}s{\ell}{\ell}$ decays,''
Phys.\ Rev.\ D {\bf 96} (2017) no.9,  095034
doi:10.1103/PhysRevD.96.095034
[arXiv:1705.06274 [hep-ph]].
%%CITATION = doi:10.1103/PhysRevD.96.095034;%%
%18 citations counted in INSPIRE as of 02 Feb 2018


%\cite{Ciuchini:2017mik}
\bibitem{Ciuchini:2017mik}
M.~Ciuchini, A.~M.~Coutinho, M.~Fedele, E.~Franco, A.~Paul, L.~Silvestrini and M.~Valli,
%``On Flavourful Easter eggs for New Physics hunger and Lepton Flavour Universality violation,''
Eur.\ Phys.\ J.\ C {\bf 77} (2017) no.10,  688
doi:10.1140/epjc/s10052-017-5270-2
[arXiv:1704.05447 [hep-ph]].
%%CITATION = doi:10.1140/epjc/s10052-017-5270-2;%%
%62 citations counted in INSPIRE as of 02 Feb 2018


%\cite{Capdevila:2017bsm}
\bibitem{Capdevila:2017bsm}
B.~Capdevila, A.~Crivellin, S.~Descotes-Genon, J.~Matias and J.~Virto,
%``Patterns of New Physics in $b\to s\ell^+\ell^-$ transitions in the light of recent data,''
JHEP {\bf 1801} (2018) 093
doi:10.1007/JHEP01(2018)093
[arXiv:1704.05340 [hep-ph]].
%%CITATION = doi:10.1007/JHEP01(2018)093;%%
%96 citations counted in INSPIRE as of 02 Feb 2018


%\cite{Lees:2012xj}
\bibitem{Lees:2012xj}
J.~P.~Lees {\it et al.} [BaBar Collaboration],
%``Evidence for an excess of $\bar{B} \to D^{(*)} \tau^-\bar{\nu}_\tau$ decays,''
Phys.\ Rev.\ Lett.\  {\bf 109} (2012) 101802
doi:10.1103/PhysRevLett.109.101802
[arXiv:1205.5442 [hep-ex]].
%%CITATION = doi:10.1103/PhysRevLett.109.101802;%%
%484 citations counted in INSPIRE as of 02 Feb 2018


%\cite{Huschle:2015rga}
\bibitem{Huschle:2015rga}
M.~Huschle {\it et al.} [Belle Collaboration],
%``Measurement of the branching ratio of $\bar{B} \to D^{(\ast)} \tau^- \bar{\nu}_\tau$ relative to $\bar{B} \to D^{(\ast)} \ell^- \bar{\nu}_\ell$ decays with hadronic tagging at Belle,''
Phys.\ Rev.\ D {\bf 92} (2015) no.7,  072014
doi:10.1103/PhysRevD.92.072014
[arXiv:1507.03233 [hep-ex]].
%%CITATION = doi:10.1103/PhysRevD.92.072014;%%
%276 citations counted in INSPIRE as of 02 Feb 2018


%\cite{Hirose:2017dxl}
\bibitem{Hirose:2017dxl}
S.~Hirose {\it et al.} [Belle Collaboration],
%``Measurement of the $\tau$ lepton polarization and $R(D^*)$ in the decay $\bar{B} \rightarrow D^* \tau^- \bar{\nu}_\tau$ with one-prong hadronic $\tau$ decays at Belle,''
Phys.\ Rev.\ D {\bf 97} (2018) no.1,  012004
doi:10.1103/PhysRevD.97.012004
[arXiv:1709.00129 [hep-ex]].
%%CITATION = doi:10.1103/PhysRevD.97.012004;%%
%14 citations counted in INSPIRE as of 02 Feb 2018


%\cite{Aaij:2015yra}
\bibitem{Aaij:2015yra}
R.~Aaij {\it et al.} [LHCb Collaboration],
%``Measurement of the ratio of branching fractions $\mathcal{B}(\bar{B}^0 \to D^{*+}\tau^{-}\bar{\nu}_{\tau})/\mathcal{B}(\bar{B}^0 \to D^{*+}\mu^{-}\bar{\nu}_{\mu})$,''
Phys.\ Rev.\ Lett.\  {\bf 115} (2015) no.11,  111803
Erratum: [Phys.\ Rev.\ Lett.\  {\bf 115} (2015) no.15,  159901]
doi:10.1103/PhysRevLett.115.159901, 10.1103/PhysRevLett.115.111803
[arXiv:1506.08614 [hep-ex]].
%%CITATION = doi:10.1103/PhysRevLett.115.159901, 10.1103/PhysRevLett.115.111803;%%
%335 citations counted in INSPIRE as of 02 Feb 2018


%\cite{Betti:2017xri}
\bibitem{Betti:2017xri}
F.~Betti [LHCb Collaboration],
%``Measurement of $\mathcal{R}(D^*)$ with Three-Prong $\tau$ Decays at LHCb,''
arXiv:1705.10651 [hep-ex].
%%CITATION = ARXIV:1705.10651;%%
%3 citations counted in INSPIRE as of 02 Feb 2018


%\cite{Amhis:2016xyh}
\bibitem{Amhis:2016xyh}
Y.~Amhis {\it et al.} [HFLAV Collaboration],
%``Averages of $b$-hadron, $c$-hadron, and $\tau$-lepton properties as of summer 2016,''
Eur.\ Phys.\ J.\ C {\bf 77} (2017) no.12,  895
doi:10.1140/epjc/s10052-017-5058-4
[arXiv:1612.07233 [hep-ex]].
%%CITATION = doi:10.1140/epjc/s10052-017-5058-4;%%
%239 citations counted in INSPIRE as of 02 Feb 2018


%\cite{Aaij:2017tyk}
\bibitem{Aaij:2017tyk}
R.~Aaij {\it et al.} [LHCb Collaboration],
%``Measurement of the ratio of branching fractions $\mathcal{B}(B_c^+\,\to\,J/\psi\tau^+\nu_\tau)$/$\mathcal{B}(B_c^+\,\to\,J/\psi\mu^+\nu_\mu)$,''
arXiv:1711.05623 [hep-ex].
%%CITATION = ARXIV:1711.05623;%%
%20 citations counted in INSPIRE as of 02 Feb 2018


%\cite{Bennett:2006fi}
\bibitem{Bennett:2006fi}
G.~W.~Bennett {\it et al.} [Muon g-2 Collaboration],
%``Final Report of the Muon E821 Anomalous Magnetic Moment Measurement at BNL,''
Phys.\ Rev.\ D {\bf 73} (2006) 072003
doi:10.1103/PhysRevD.73.072003
[hep-ex/0602035].
%%CITATION = doi:10.1103/PhysRevD.73.072003;%%
%1536 citations counted in INSPIRE as of 02 Feb 2018


%\cite{Nyffeler:2016gnb}
\bibitem{Nyffeler:2016gnb}
A.~Nyffeler,
%``Precision of a data-driven estimate of hadronic light-by-light scattering in the muon $g-2$: Pseudoscalar-pole contribution,''
Phys.\ Rev.\ D {\bf 94} (2016) no.5,  053006
doi:10.1103/PhysRevD.94.053006
[arXiv:1602.03398 [hep-ph]].
%%CITATION = doi:10.1103/PhysRevD.94.053006;%%
%35 citations counted in INSPIRE as of 02 Feb 2018


%\cite{Jegerlehner:2009ry}
\bibitem{Jegerlehner:2009ry}
F.~Jegerlehner and A.~Nyffeler,
%``The Muon g-2,''
Phys.\ Rept.\  {\bf 477} (2009) 1
doi:10.1016/j.physrep.2009.04.003
[arXiv:0902.3360 [hep-ph]].
%%CITATION = doi:10.1016/j.physrep.2009.04.003;%%
%733 citations counted in INSPIRE as of 02 Feb 2018


%\cite{Djouadi:1989md}
\bibitem{Djouadi:1989md}
A.~Djouadi, T.~Kohler, M.~Spira and J.~Tutas,
%``(e b), (e t) TYPE LEPTOQUARKS AT e p COLLIDERS,''
Z.\ Phys.\ C {\bf 46} (1990) 679.
doi:10.1007/BF01560270
%%CITATION = doi:10.1007/BF01560270;%%
%59 citations counted in INSPIRE as of 02 Feb 2018


%\cite{Davidson:1993qk}
\bibitem{Davidson:1993qk}
S.~Davidson, D.~C.~Bailey and B.~A.~Campbell,
%``Model independent constraints on leptoquarks from rare processes,''
Z.\ Phys.\ C {\bf 61} (1994) 613
doi:10.1007/BF01552629
[hep-ph/9309310].
%%CITATION = doi:10.1007/BF01552629;%%
%403 citations counted in INSPIRE as of 02 Feb 2018


%\cite{Couture:1995he}
\bibitem{Couture:1995he}
G.~Couture and H.~Konig,
%``Bounds on second generation scalar leptoquarks from the anomalous magnetic moment of the muon,''
Phys.\ Rev.\ D {\bf 53} (1996) 555
doi:10.1103/PhysRevD.53.555
[hep-ph/9507263].
%%CITATION = doi:10.1103/PhysRevD.53.555;%%
%31 citations counted in INSPIRE as of 02 Feb 2018


%\cite{Bauer:2015knc}
\bibitem{Bauer:2015knc}
M.~Bauer and M.~Neubert,
%``Minimal Leptoquark Explanation for the R$_{D^{(*)}}$ , R$_K$ , and $(g-2)_g$ Anomalies,''
Phys.\ Rev.\ Lett.\  {\bf 116} (2016) no.14,  141802
doi:10.1103/PhysRevLett.116.141802
[arXiv:1511.01900 [hep-ph]].
%%CITATION = doi:10.1103/PhysRevLett.116.141802;%%
%138 citations counted in INSPIRE as of 02 Feb 2018


%\cite{Das:2016vkr}
\bibitem{Das:2016vkr}
D.~Das, C.~Hati, G.~Kumar and N.~Mahajan,
%``Towards a unified explanation of $R_{D^{(\ast)}}$, $R_{K}$ and $(g-2)_{\mu}$ anomalies in a left-right model with leptoquarks,''
Phys.\ Rev.\ D {\bf 94} (2016) 055034
doi:10.1103/PhysRevD.94.055034
[arXiv:1605.06313 [hep-ph]].
%%CITATION = doi:10.1103/PhysRevD.94.055034;%%
%60 citations counted in INSPIRE as of 02 Feb 2018


%\cite{Buchmuller:1986zs}
\bibitem{Buchmuller:1986zs}
W.~Buchmuller, R.~Ruckl and D.~Wyler,
%``Leptoquarks in Lepton - Quark Collisions,''
Phys.\ Lett.\ B {\bf 191} (1987) 442
Erratum: [Phys.\ Lett.\ B {\bf 448} (1999) 320].
doi:10.1016/S0370-2693(99)00014-3, 10.1016/0370-2693(87)90637-X
%%CITATION = doi:10.1016/S0370-2693(99)00014-3, 10.1016/0370-2693(87)90637-X;%%
%648 citations counted in INSPIRE as of 02 Feb 2018


%\cite{Baer:2013cma}
\bibitem{Baer:2013cma}
H.~Baer {\it et al.},
%``The International Linear Collider Technical Design Report - Volume 2: Physics,''
arXiv:1306.6352 [hep-ph].
%%CITATION = ARXIV:1306.6352;%%
%555 citations counted in INSPIRE as of 02 Feb 2018


%\cite{Gomez-Ceballos:2013zzn}
\bibitem{Gomez-Ceballos:2013zzn}
M.~Bicer {\it et al.} [TLEP Design Study Working Group],
%``First Look at the Physics Case of TLEP,''
JHEP {\bf 1401} (2014) 164
doi:10.1007/JHEP01(2014)164
[arXiv:1308.6176 [hep-ex]].
%%CITATION = doi:10.1007/JHEP01(2014)164;%%
%385 citations counted in INSPIRE as of 02 Feb 2018


%\cite{ColuccioLeskow:2016dox}
\bibitem{ColuccioLeskow:2016dox}
E.~Coluccio Leskow, G.~D'Ambrosio, A.~Crivellin and D.~Müller,
%``$(g-2)\mu$, lepton flavor violation, and $Z$ decays with leptoquarks: Correlations and future prospects,''
Phys.\ Rev.\ D {\bf 95} (2017) no.5,  055018
doi:10.1103/PhysRevD.95.055018
[arXiv:1612.06858 [hep-ph]].
%%CITATION = doi:10.1103/PhysRevD.95.055018;%%
%15 citations counted in INSPIRE as of 02 Feb 2018


%\cite{Gripaios:2014tna}
\bibitem{Gripaios:2014tna}
B.~Gripaios, M.~Nardecchia and S.~A.~Renner,
%``Composite leptoquarks and anomalies in $B$-meson decays,''
JHEP {\bf 1505} (2015) 006
doi:10.1007/JHEP05(2015)006
[arXiv:1412.1791 [hep-ph]].
%%CITATION = doi:10.1007/JHEP05(2015)006;%%
%131 citations counted in INSPIRE as of 02 Feb 2018


%\cite{Fajfer:2015ycq}
\bibitem{Fajfer:2015ycq}
S.~Fajfer and N.~Košnik,
%``Vector leptoquark resolution of $R_K$ and $R_{D^{(*)}}$ puzzles,''
Phys.\ Lett.\ B {\bf 755} (2016) 270
doi:10.1016/j.physletb.2016.02.018
[arXiv:1511.06024 [hep-ph]].
%%CITATION = doi:10.1016/j.physletb.2016.02.018;%%
%114 citations counted in INSPIRE as of 02 Feb 2018


%\cite{Becirevic:2015asa}
\bibitem{Becirevic:2015asa}
D.~Bečirević, S.~Fajfer and N.~Košnik,
%``Lepton flavor nonuniversality in b→sℓ$^+$ℓ$^-$ processes,''
Phys.\ Rev.\ D {\bf 92} (2015) no.1,  014016
doi:10.1103/PhysRevD.92.014016
[arXiv:1503.09024 [hep-ph]].
%%CITATION = doi:10.1103/PhysRevD.92.014016;%%
%101 citations counted in INSPIRE as of 02 Feb 2018


%\cite{Greljo:2015mma}
\bibitem{Greljo:2015mma}
A.~Greljo, G.~Isidori and D.~Marzocca,
%``On the breaking of Lepton Flavor Universality in B decays,''
JHEP {\bf 1507} (2015) 142
doi:10.1007/JHEP07(2015)142
[arXiv:1506.01705 [hep-ph]].
%%CITATION = doi:10.1007/JHEP07(2015)142;%%
%123 citations counted in INSPIRE as of 02 Feb 2018


%\cite{Calibbi:2015kma}
\bibitem{Calibbi:2015kma}
L.~Calibbi, A.~Crivellin and T.~Ota,
%``Effective Field Theory Approach to b→sℓℓ(′), B→K(*)ν$\overline{ν}$ and B→D(*)τν with Third Generation Couplings,''
Phys.\ Rev.\ Lett.\  {\bf 115} (2015) 181801
doi:10.1103/PhysRevLett.115.181801
[arXiv:1506.02661 [hep-ph]].
%%CITATION = doi:10.1103/PhysRevLett.115.181801;%%
%136 citations counted in INSPIRE as of 02 Feb 2018


%\cite{Alonso:2015sja}
\bibitem{Alonso:2015sja}
R.~Alonso, B.~Grinstein and J.~Martin Camalich,
%``Lepton universality violation and lepton flavor conservation in $B$-meson decays,''
JHEP {\bf 1510} (2015) 184
doi:10.1007/JHEP10(2015)184
[arXiv:1505.05164 [hep-ph]].
%%CITATION = doi:10.1007/JHEP10(2015)184;%%
%145 citations counted in INSPIRE as of 02 Feb 2018


%\cite{Barbieri:2015yvd}
\bibitem{Barbieri:2015yvd}
R.~Barbieri, G.~Isidori, A.~Pattori and F.~Senia,
%``Anomalies in $B$-decays and $U(2)$ flavour symmetry,''
Eur.\ Phys.\ J.\ C {\bf 76} (2016) no.2,  67
doi:10.1140/epjc/s10052-016-3905-3
[arXiv:1512.01560 [hep-ph]].
%%CITATION = doi:10.1140/epjc/s10052-016-3905-3;%%
%95 citations counted in INSPIRE as of 02 Feb 2018


%\cite{Becirevic:2016yqi}
\bibitem{Becirevic:2016yqi}
D.~Bečirević, S.~Fajfer, N.~Košnik and O.~Sumensari,
%``Leptoquark model to explain the $B$-physics anomalies, $R_K$ and $R_D$,''
Phys.\ Rev.\ D {\bf 94} (2016) no.11,  115021
doi:10.1103/PhysRevD.94.115021
[arXiv:1608.08501 [hep-ph]].
%%CITATION = doi:10.1103/PhysRevD.94.115021;%%
%88 citations counted in INSPIRE as of 02 Feb 2018


%\cite{Becirevic:2016oho}
\bibitem{Becirevic:2016oho}
D.~Bečirević, N.~Košnik, O.~Sumensari and R.~Zukanovich Funchal,
%``Palatable Leptoquark Scenarios for Lepton Flavor Violation in Exclusive $b\to s\ell_1\ell_2$ modes,''
JHEP {\bf 1611} (2016) 035
doi:10.1007/JHEP11(2016)035
[arXiv:1608.07583 [hep-ph]].
%%CITATION = doi:10.1007/JHEP11(2016)035;%%
%49 citations counted in INSPIRE as of 02 Feb 2018


%\cite{Becirevic:2017jtw}
\bibitem{Becirevic:2017jtw}
D.~Bečirević and O.~Sumensari,
%``A leptoquark model to accommodate $R_K^\mathrm{exp} < R_K^\mathrm{SM}$ and $R_{K^\ast}^\mathrm{exp} < R_{K^\ast}^\mathrm{SM}$,''
JHEP {\bf 1708} (2017) 104
doi:10.1007/JHEP08(2017)104
[arXiv:1704.05835 [hep-ph]].
%%CITATION = doi:10.1007/JHEP08(2017)104;%%
%38 citations counted in INSPIRE as of 02 Feb 2018


%\cite{Aloni:2017ixa}
\bibitem{Aloni:2017ixa}
D.~Aloni, A.~Dery, C.~Frugiuele and Y.~Nir,
%``Testing minimal flavor violation in leptoquark models of the $ {R_K}_{{}^{\left(\ast \right)}} $ anomaly,''
JHEP {\bf 1711} (2017) 109
doi:10.1007/JHEP11(2017)109
[arXiv:1708.06161 [hep-ph]].
%%CITATION = doi:10.1007/JHEP11(2017)109;%%
%5 citations counted in INSPIRE as of 02 Feb 2018


%\cite{Chauhan:2017ndd}
\bibitem{Chauhan:2017ndd}
B.~Chauhan, B.~Kindra and A.~Narang,
%``A Leptoquark explanation for $(g-2)_\mu$, $R_K$, $R_{K^*}$ and, IceCube PeV events,''
arXiv:1706.04598 [hep-ph].
%%CITATION = ARXIV:1706.04598;%%
%9 citations counted in INSPIRE as of 02 Feb 2018


%\cite{Sumensari:2017mud}
\bibitem{Sumensari:2017mud}
O.~Sumensari,
%``Leptoquark models for the $B$-physics anomalies,''
arXiv:1705.07591 [hep-ph].
%%CITATION = ARXIV:1705.07591;%%
%3 citations counted in INSPIRE as of 02 Feb 2018


%\cite{Chen:2017hir}
\bibitem{Chen:2017hir}
C.~H.~Chen, T.~Nomura and H.~Okada,
%``Excesses of muon $g-2$, $R_{D^{(\ast)}}$, and $R_K$ in a leptoquark model,''
Phys.\ Lett.\ B {\bf 774} (2017) 456
doi:10.1016/j.physletb.2017.10.005
[arXiv:1703.03251 [hep-ph]].
%%CITATION = doi:10.1016/j.physletb.2017.10.005;%%
%24 citations counted in INSPIRE as of 02 Feb 2018


%\cite{Crivellin:2017dsk}
\bibitem{Crivellin:2017dsk}
A.~Crivellin, D.~Mueller, A.~Signer and Y.~Ulrich,
%``Correlating Lepton Flavour (Universality) Violation in $B$ Decays with $\mu\to e\gamma$ using Leptoquarks,''
arXiv:1706.08511 [hep-ph].
%%CITATION = ARXIV:1706.08511;%%
%11 citations counted in INSPIRE as of 02 Feb 2018


%\cite{TheMEG:2016wtm}
\bibitem{TheMEG:2016wtm}
A.~M.~Baldini {\it et al.} [MEG Collaboration],
%``Search for the lepton flavour violating decay $\mu ^+ \rightarrow \mathrm {e}^+ \gamma $ with the full dataset of the MEG experiment,''
Eur.\ Phys.\ J.\ C {\bf 76} (2016) no.8,  434
doi:10.1140/epjc/s10052-016-4271-x
[arXiv:1605.05081 [hep-ex]].
%%CITATION = doi:10.1140/epjc/s10052-016-4271-x;%%
%163 citations counted in INSPIRE as of 02 Feb 2018


%\cite{Crivellin:2012ye}
\bibitem{Crivellin:2012ye}
A.~Crivellin, C.~Greub and A.~Kokulu,
%``Explaining $B\to D\tau\nu$, $B\to D^*\tau\nu$ and $B\to \tau\nu$ in a 2HDM of type III,''
Phys.\ Rev.\ D {\bf 86} (2012) 054014
doi:10.1103/PhysRevD.86.054014
[arXiv:1206.2634 [hep-ph]].
%%CITATION = doi:10.1103/PhysRevD.86.054014;%%
%194 citations counted in INSPIRE as of 02 Feb 2018


%\cite{Tanaka:2012nw}
\bibitem{Tanaka:2012nw}
M.~Tanaka and R.~Watanabe,
%``New physics in the weak interaction of $\bar B\to D^{(*)}\tau\bar\nu$,''
Phys.\ Rev.\ D {\bf 87} (2013) no.3,  034028
doi:10.1103/PhysRevD.87.034028
[arXiv:1212.1878 [hep-ph]].
%%CITATION = doi:10.1103/PhysRevD.87.034028;%%
%130 citations counted in INSPIRE as of 02 Feb 2018


%\cite{Celis:2012dk}
\bibitem{Celis:2012dk}
A.~Celis, M.~Jung, X.~Q.~Li and A.~Pich,
%``Sensitivity to charged scalars in $\boldsymbol{B\to D^{(*)}\tau\nu_\tau}$ and $\boldsymbol{B\to\tau\nu_\tau}$ decays,''
JHEP {\bf 1301} (2013) 054
doi:10.1007/JHEP01(2013)054
[arXiv:1210.8443 [hep-ph]].
%%CITATION = doi:10.1007/JHEP01(2013)054;%%
%161 citations counted in INSPIRE as of 02 Feb 2018


%\cite{Crivellin:2013wna}
\bibitem{Crivellin:2013wna}
A.~Crivellin, A.~Kokulu and C.~Greub,
%``Flavor-phenomenology of two-Higgs-doublet models with generic Yukawa structure,''
Phys.\ Rev.\ D {\bf 87} (2013) no.9,  094031
doi:10.1103/PhysRevD.87.094031
[arXiv:1303.5877 [hep-ph]].
%%CITATION = doi:10.1103/PhysRevD.87.094031;%%
%142 citations counted in INSPIRE as of 02 Feb 2018


%\cite{Crivellin:2015hha}
\bibitem{Crivellin:2015hha}
A.~Crivellin, J.~Heeck and P.~Stoffer,
%``A perturbed lepton-specific two-Higgs-doublet model facing experimental hints for physics beyond the Standard Model,''
Phys.\ Rev.\ Lett.\  {\bf 116} (2016) no.8,  081801
doi:10.1103/PhysRevLett.116.081801
[arXiv:1507.07567 [hep-ph]].
%%CITATION = doi:10.1103/PhysRevLett.116.081801;%%
%92 citations counted in INSPIRE as of 02 Feb 2018


%\cite{Barbieri:2016las}
\bibitem{Barbieri:2016las}
R.~Barbieri, C.~W.~Murphy and F.~Senia,
%``B-decay Anomalies in a Composite Leptoquark Model,''
Eur.\ Phys.\ J.\ C {\bf 77} (2017) no.1,  8
doi:10.1140/epjc/s10052-016-4578-7
[arXiv:1611.04930 [hep-ph]].
%%CITATION = doi:10.1140/epjc/s10052-016-4578-7;%%
%45 citations counted in INSPIRE as of 02 Feb 2018


%\cite{Chauhan:2017uil}
\bibitem{Chauhan:2017uil}
B.~Chauhan and B.~Kindra,
%``Invoking Chiral Vector Leptoquark to explain LFU violation in B Decays,''
arXiv:1709.09989 [hep-ph].
%%CITATION = ARXIV:1709.09989;%%
%6 citations counted in INSPIRE as of 02 Feb 2018


%\cite{Dorsner:2017ufx}
\bibitem{Dorsner:2017ufx}
I.~Doršner, S.~Fajfer, D.~A.~Faroughy and N.~Košnik,
%``The role of the $S_3$ GUT leptoquark in flavor universality and collider searches,''
JHEP {\bf 1710} (2017) 188
doi:10.1007/JHEP10(2017)188
[arXiv:1706.07779 [hep-ph]].
%%CITATION = doi:10.1007/JHEP10(2017)188;%%
%18 citations counted in INSPIRE as of 02 Feb 2018


%\cite{Popov:2016fzr}
\bibitem{Popov:2016fzr}
O.~Popov and G.~A.~White,
%``One Leptoquark to unify them? Neutrino masses and unification in the light of $(g-2)_\mu$, $R_{D^{(\star)}}$ and $R_K$ anomalies,''
Nucl.\ Phys.\ B {\bf 923} (2017) 324
doi:10.1016/j.nuclphysb.2017.08.007
[arXiv:1611.04566 [hep-ph]].
%%CITATION = doi:10.1016/j.nuclphysb.2017.08.007;%%
%9 citations counted in INSPIRE as of 02 Feb 2018


%\cite{Hiller:2016kry}
\bibitem{Hiller:2016kry}
G.~Hiller, D.~Loose and K.~Schönwald,
%``Leptoquark Flavor Patterns & B Decay Anomalies,''
JHEP {\bf 1612} (2016) 027
doi:10.1007/JHEP12(2016)027
[arXiv:1609.08895 [hep-ph]].
%%CITATION = doi:10.1007/JHEP12(2016)027;%%
%40 citations counted in INSPIRE as of 02 Feb 2018


%\cite{Sahoo:2016pet}
\bibitem{Sahoo:2016pet}
S.~Sahoo, R.~Mohanta and A.~K.~Giri,
%``Explaining the $R_{K}$ and $R_{D^{(*)}}$ anomalies with vector leptoquarks,''
Phys.\ Rev.\ D {\bf 95} (2017) no.3,  035027
doi:10.1103/PhysRevD.95.035027
[arXiv:1609.04367 [hep-ph]].
%%CITATION = doi:10.1103/PhysRevD.95.035027;%%
%52 citations counted in INSPIRE as of 02 Feb 2018


%\cite{Altmannshofer:2017poe}
\bibitem{Altmannshofer:2017poe}
W.~Altmannshofer, P.~S.~Bhupal Dev and A.~Soni,
%``$R_{D^{(*)}}$ anomaly: A possible hint for natural supersymmetry with $R$-parity violation,''
Phys.\ Rev.\ D {\bf 96} (2017) no.9,  095010
doi:10.1103/PhysRevD.96.095010
[arXiv:1704.06659 [hep-ph]].
%%CITATION = doi:10.1103/PhysRevD.96.095010;%%
%17 citations counted in INSPIRE as of 02 Feb 2018


%\cite{Akeroyd:2017mhr}
\bibitem{Akeroyd:2017mhr}
A.~G.~Akeroyd and C.~H.~Chen,
%``Constraint on the branching ratio of $B_c \to \tau \bar{\nu}$ from LEP1 and consequences for $R(D^{(*)})$ anomaly,''
Phys.\ Rev.\ D {\bf 96} (2017) no.7,  075011
doi:10.1103/PhysRevD.96.075011
[arXiv:1708.04072 [hep-ph]].
%%CITATION = doi:10.1103/PhysRevD.96.075011;%%
%13 citations counted in INSPIRE as of 02 Feb 2018


%\cite{Freytsis:2015qca}
\bibitem{Freytsis:2015qca}
M.~Freytsis, Z.~Ligeti and J.~T.~Ruderman,
%``Flavor models for $\bar{B} \to D^{(*)} \tau \bar{\nu}$,''
Phys.\ Rev.\ D {\bf 92} (2015) no.5,  054018
doi:10.1103/PhysRevD.92.054018
[arXiv:1506.08896 [hep-ph]].
%%CITATION = doi:10.1103/PhysRevD.92.054018;%%
%118 citations counted in INSPIRE as of 02 Feb 2018


%\cite{Celis:2016azn}
\bibitem{Celis:2016azn}
A.~Celis, M.~Jung, X.~Q.~Li and A.~Pich,
%``Scalar contributions to $b\to c (u) \tau \nu$ transitions,''
Phys.\ Lett.\ B {\bf 771} (2017) 168
doi:10.1016/j.physletb.2017.05.037
[arXiv:1612.07757 [hep-ph]].
%%CITATION = doi:10.1016/j.physletb.2017.05.037;%%
%37 citations counted in INSPIRE as of 02 Feb 2018


%\cite{Li:2016vvp}
\bibitem{Li:2016vvp}
X.~Q.~Li, Y.~D.~Yang and X.~Zhang,
%``Revisiting the one leptoquark solution to the R(D$^{(∗)}$) anomalies and its phenomenological implications,''
JHEP {\bf 1608} (2016) 054
doi:10.1007/JHEP08(2016)054
[arXiv:1605.09308 [hep-ph]].
%%CITATION = doi:10.1007/JHEP08(2016)054;%%
%55 citations counted in INSPIRE as of 02 Feb 2018


%\cite{Faroughy:2016osc}
\bibitem{Faroughy:2016osc}
D.~A.~Faroughy, A.~Greljo and J.~F.~Kamenik,
%``Confronting lepton flavor universality violation in B decays with high-$p_T$ tau lepton searches at LHC,''
Phys.\ Lett.\ B {\bf 764} (2017) 126
doi:10.1016/j.physletb.2016.11.011
[arXiv:1609.07138 [hep-ph]].
%%CITATION = doi:10.1016/j.physletb.2016.11.011;%%
%52 citations counted in INSPIRE as of 02 Feb 2018


%\cite{Capdevila:2017iqn}
\bibitem{Capdevila:2017iqn}
B.~Capdevila, A.~Crivellin, S.~Descotes-Genon, L.~Hofer and J.~Matias,
%``Searching for New Physics with $b\to s\tau^+\tau^-$ processes,''
arXiv:1712.01919 [hep-ph].
%%CITATION = ARXIV:1712.01919;%%
%2 citations counted in INSPIRE as of 02 Feb 2018


%\cite{Crivellin:2017zlb}
\bibitem{Crivellin:2017zlb}
A.~Crivellin, D.~Müller and T.~Ota,
%``Simultaneous explanation of R(D$^{(∗)}$) and b→sμ$^{+}$ μ$^{−}$: the last scalar leptoquarks standing,''
JHEP {\bf 1709} (2017) 040
doi:10.1007/JHEP09(2017)040
[arXiv:1703.09226 [hep-ph]].
%%CITATION = doi:10.1007/JHEP09(2017)040;%%
%48 citations counted in INSPIRE as of 02 Feb 2018


%\cite{Aaij:2017xqt}
\bibitem{Aaij:2017xqt}
R.~Aaij {\it et al.} [LHCb Collaboration],
%``Search for the decays $B_s^0\to\tau^+\tau^-$ and $B^0\to\tau^+\tau^-$,''
Phys.\ Rev.\ Lett.\  {\bf 118} (2017) no.25,  251802
doi:10.1103/PhysRevLett.118.251802
[arXiv:1703.02508 [hep-ex]].
%%CITATION = doi:10.1103/PhysRevLett.118.251802;%%
%24 citations counted in INSPIRE as of 02 Feb 2018


%\cite{DiLuzio:2017vat}
\bibitem{DiLuzio:2017vat}
L.~Di Luzio, A.~Greljo and M.~Nardecchia,
%``Gauge leptoquark as the origin of B-physics anomalies,''
Phys.\ Rev.\ D {\bf 96} (2017) no.11,  115011
doi:10.1103/PhysRevD.96.115011
[arXiv:1708.08450 [hep-ph]].
%%CITATION = doi:10.1103/PhysRevD.96.115011;%%
%15 citations counted in INSPIRE as of 02 Feb 2018


%\cite{Calibbi:2017qbu}
\bibitem{Calibbi:2017qbu}
L.~Calibbi, A.~Crivellin and T.~Li,
%``A model of vector leptoquarks in view of the $B$-physics anomalies,''
arXiv:1709.00692 [hep-ph].
%%CITATION = ARXIV:1709.00692;%%
%17 citations counted in INSPIRE as of 02 Feb 2018


%\cite{Bordone:2017bld}
\bibitem{Bordone:2017bld}
M.~Bordone, C.~Cornella, J.~Fuentes-Martin and G.~Isidori,
%``A three-site gauge model for flavor hierarchies and flavor anomalies,''
arXiv:1712.01368 [hep-ph].
%%CITATION = ARXIV:1712.01368;%%
%7 citations counted in INSPIRE as of 02 Feb 2018


%\cite{Barbieri:2017tuq}
\bibitem{Barbieri:2017tuq}
R.~Barbieri and A.~Tesi,
%``B-decay anomalies in Pati-Salam SU(4),''
arXiv:1712.06844 [hep-ph].
%%CITATION = ARXIV:1712.06844;%%
%5 citations counted in INSPIRE as of 02 Feb 2018


\end{thebibliography}

\end{document}